\documentclass[useAMS,usenatbib,usegraphicx]{mn2e}
\include{epsf}

\title[AGN Heating in Elliptical Galaxies]
{Self-Regulated AGN Heating in Elliptical Galaxies}
\author[D.\ Kawata and B.K.\ Gibson]
 {D.~Kawata and B.K.~Gibson
\thanks{E-mail: dkawata,bgibson@astro.swin.edu.au}
\\
Centre for Astrophysics and Supercomputing, 
 Swinburne University of Technology, Hawthorn VIC 3122, Australia
}
\date{Accepted .
      Received ;
      in original form }

\pagerange{\pageref{firstpage}--\pageref{lastpage}}
\pubyear{2004}

\begin{document}

\maketitle

\label{firstpage}

\begin{abstract}
We study the effect of active galactic nuclei (AGN) heating on the
chemodynamical evolution of elliptical galaxies and their X-ray and
optical properties using high-resolution $\Lambda$-dominated cold dark
matter cosmological simulations. Our model considers an AGN as being
``active'' when a convergent gas inflow condition exists within a galaxy's
nucleus; otherwise, the AGN is assumed to remain dormant. This induces a
self-regulated activity for the AGN, the result of which leads to a stable
hot corona and the suppression of significant late-time star formation -
characteristics not encountered in traditional chemodynamical models of
ellipticals. These properties of our AGN heating model leads to a system
consistent with both the X-ray and optical properties of comparable
elliptical galaxies. 
\end{abstract}

\begin{keywords}
galaxies: elliptical and lenticular, cD --- galaxies:
formation---galaxies: evolution --- galaxies: stellar content
\end{keywords}

\section{Introduction}
\label{intro-sec}

One of the primary science drivers for the next generation of ground- and
space-based observational facilities
which spread over multi-wavelength
is an understanding of the physics of
galaxy formation and evolution.  In the case of elliptical galaxies, the
optical regime provides constraints on the properties of the underlying
stellar populations, while the X-ray regime yields insights into the
physical conditions of the associated hot interstellar (coronal) medium. 

\citet[][KG03b]{kg03b} presented a first attempt at explaining both the
X-ray and optical properties of ellipticals via the use of self-consistent
cosmological simulations. Using a standard ``recipe'' for galaxy
formation, they found that radiative cooling is important to interpret the
observed X-ray luminosity (${\rm L_X}$), temperature (${\rm T_X}$), and
metallicity (${\rm [Fe/H]_X}$) of the hot gas of elliptical galaxies.
However, these models were subject to an unavoidable serious problem, in
that the cooled gas unavoidably led to excessive star formation at low
redshift \citep[see also][]{so98,lbk00,ktt03,tbs03,mnse03,bms04}.  This
late-time star formation led to associated stellar populations which were
too blue with respect to observations. 

KG03b suggested that a heating mechanism not included in their simulations
would be required to suppress this enhanced cooling and consequent star
formation. In this paper, we examine one such potential mechanism -
heating by an active galactic nucleus (AGN). 
In such a picture, one might expect an AGN could become ``active'' only
when being fed ``fuel'' - i.e., when cold gas inflows into the central
region, the AGN would activate and heat the surrounding gas, and
potentially balance the associated radiative cooling
\citep[e.g.][]{csfb02,pn04,gds04,so04}.
Once the infalling
fuel source is quenched by the AGN heating, the AGN reverts to its
quiescent state, and cycle can begin anew.  This self-regulation provides
heating (suppressing star formation) without compromising cooling
(required to recover the empirical X-ray properties). Recent
high-resolution X-ray images taken by {\it Chandra} reveal that the
central hot gas of ellipticals is not smoothly distributed, but possesses
cavities on scales comparable to the radio emission
\citep[e.g.][]{bvfen93,fj01,jfv02}. The X-ray holes are also seen on the
cluster scale ($>50$ kpc), and are often coincident with extended radio
lobes \citep[e.g.][]{mwn00, atn01,bsmw01,fsk02,hcrb02}. These features are
considered to be a relic of the influence of AGN activity upon the hot gas
in ellipticals and clusters \citep{cbkbf01}. 

Our current study examines the effect of AGN heating on both the optical
\it and \rm X-ray properties of giant elliptical galaxies in the context
of a self-consistent treatment of cosmological evolution. Several earlier
studies of the effect of AGN heating on the hot gas of ellipticals can be
seen in \citet{bt95,cbkbf01,co01,qbb01,bm02,bk02,sbs02,ba03,obbs04,hb04}. 
While all important contributions to the field, these previous studies
were (a) restricted in their temporal coverage to $<$15\% of a Hubble
time, and (b) did not trace star formation or mass accretion
self-consistently within a $\Lambda$-dominated cold dark matter
($\Lambda$CDM) cosmology. Our cosmological simulation makes it possible to
investigate whether or not AGN heating can lead to (a) a stable condition
for the hot gas, and (b) suppress late-time star formation. 

\begin{figure}
\centering
\includegraphics[width=\hsize]{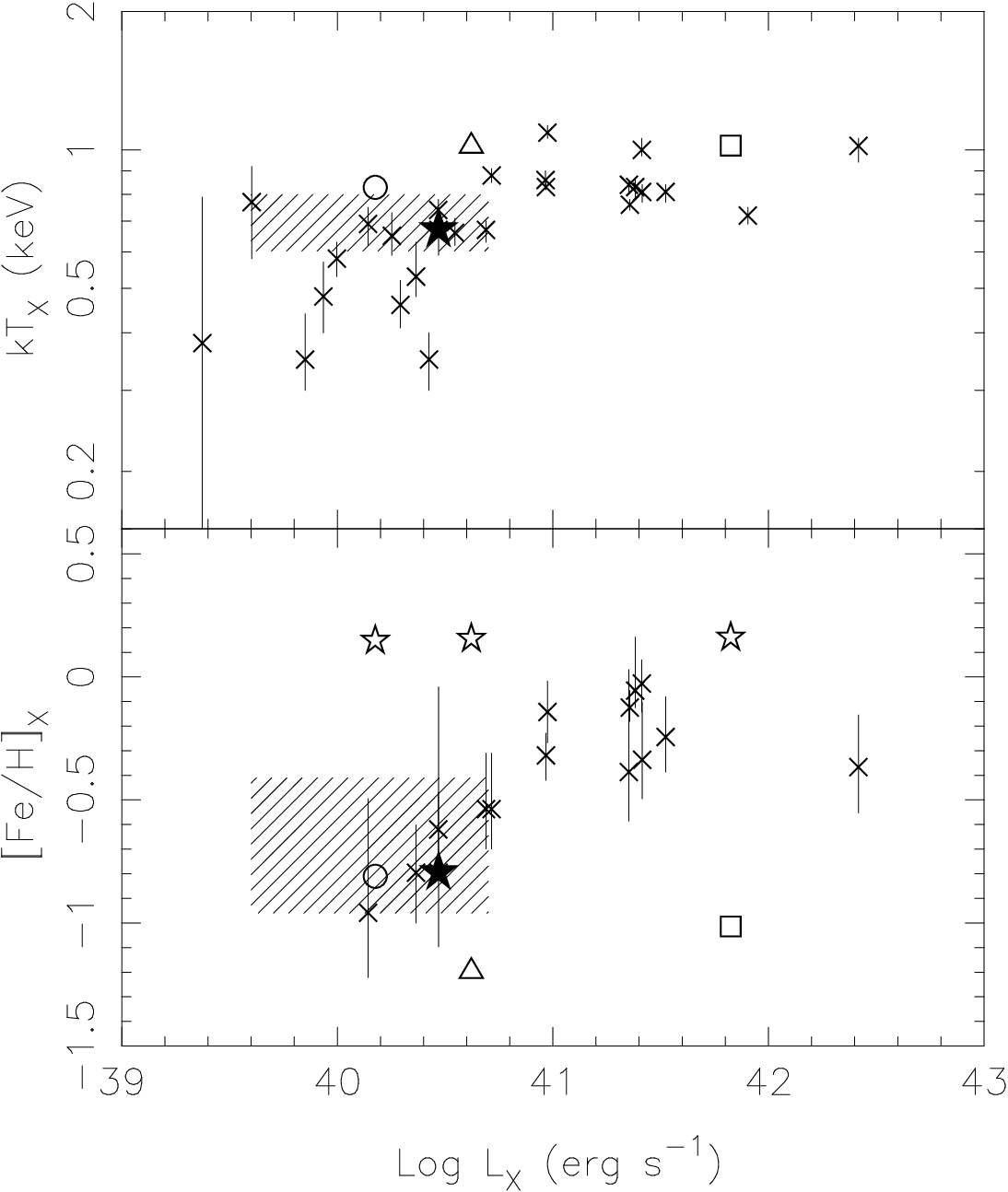}
\caption{Comparison of the simulated and observed (crosses with error 
bars) ${\rm L_X-T_X}$ relations (upper) and ${\rm [Fe/H]_X-L_X}$ relations
(lower). The square/circle/triangle indicates the predictions of 
Model 1/2/2L. 
Open stars in the lower panel show the predicted mean metallicity of
the stellar component for the two models. The observational data are from
MOM00. 
The shaded area represents the range of ${\rm L_X}$
and ${\rm T_X}$ for a sample of elliptical galaxies with low-${\rm L_X}$ and
high-${\rm T_X}$ (group 5 in MOM00), and 90 \% confidence limit
of ${\rm [Fe/H]_X}$ from the composite X-ray spectrum of these galaxies.
The solid star corresponds to one of these galaxies, NGC 3923.
}
\label{lx-fig}
\end{figure}

\section{Methods}
\label{meth-sec}

Our simulations were carried out using the galactic chemodynamics
code {\tt GCD+} \citep{kg03a} with initial conditions identical to
those described in KG03b.
{\tt GCD+} is a three-dimensional tree $N$-body/smoothed
particle hydrodynamics (SPH) code which incorporates self-gravity,
hydrodynamics, radiative cooling, star formation, supernovae (SNe)
feedback, and metal enrichment. {\tt GCD+} takes account of the chemical
enrichment by both Type~II (SNe~II) and Type~Ia (SNe~Ia) SNe, mass-loss
from intermediate mass stars, and follows the chemical enrichment history
of both the stellar and gas components of the system. Here, we briefly
describe our cosmological simulation models. 

We adopt a $\Lambda$CDM cosmology ($\Omega_0 $=0.3, $\Lambda_0$=0.7,
$\Omega_{\rm b}$=0.019$h^{-2}$, $h$=0.7, and $\sigma_8$=0.9) and use a
multi-resolution technique to achieve high-resolution in the regions of
interest, including the tidal forces from neighbouring large-scale
structures. The initial conditions for the simulations are constructed
using the public software {\tt GRAFIC2} \citep{eb01}. Gas dynamics and
star formation are included only within the relevant high-resolution
region ($\sim$12~Mpc at $z$=0); the surrounding low-resolution region
($\sim$43~Mpc) contributes to the high-resolution region only through
gravity.
 Consequently, the initial condition consists of total 190093 dark matter 
particles and 134336 gas particles.
The mass and softening length of individual gas (dark matter)
particles in the high-resolution region are $5.86\times10^7$
($3.95\times10^8$) ${\rm M}_\odot$ and 2.27 (4.29) kpc, respectively.
KG03b found an appropriate elliptical galaxy analog in the high-resolution
region, which acts as the focus for this study. The total virial mass of
this galaxy is $2\times10^{13}$~${\rm M_\odot}$.
The target
galaxy is relatively isolated, with only a few low-mass satellites
remaining at $z$=0. Fig.~1 of KG03b shows the morphological evolution of
dark matter in the simulation volume and the evolution of the stellar
component of the target galaxy. The galaxy forms through conventional
hierarchical clustering between redshifts $z$=3 and $z$=1. The morphology
has not changed dramatically since $z$=1. 

KG03b presented the results of three different radiative cooling and SNe
feedback models. We use one of their models as a reference model for
comparison with this study; Model 1 here corresponds to Model~C of KG03b,
which includes cooling and strong SNe feedback.\footnote{Thermal energy of
$10^{52}$ erg is applied per supernova, admittedly unrealistic in terms of
energetics, but we are only using the model for comparison purposes.} An
additional model (Model 2) is constructed, to study the effect of heating
by an AGN \citep[see also][]{sdh04a,sdh04b}. 
In Model 2, strong thermal SNe feedback (as in Model 1) is also
adopted. In addition, the most bounded star particle in the target galaxy
at $z=1$ is assumed to be the heating source - i.e., the AGN 
``particle''.\footnote{We acknowledge that ``turning on'' AGN heating
at redshift $z$=1 ``misses'' the peak of AGN activity in the Universe by
$\sim$2~Gyr; our choice of $z$=1 for this particular model was not driven
by empirical cosmological AGN activity arguments, but by identifying 
the redshift for this model at which a stable ``nucleus'' could be
identified.  We will explore this limitation in a future study.}
Since any AGN heating necessarily requires a fuel (gas) source, we assume
that the AGN heating is active only when the divergence of the velocity
field of the gas surrounding the AGN particle is negative - i.e., $\langle
{\bf \nabla} \cdot \mbox{\boldmath$v$}\rangle<0$. The divergence of the
velocity field of the neighbour gas particles is calculated using the SPH
scheme. A constant thermal energy of $10^{44} {\rm erg\ s}^{-1}$ is
deposited to the neighbour gas particles when the above condition is
satisfied. This energy is roughly consistent with observational estimates
\citep[e.g.][]{bsm03}. The energy is deposited to the neighbour gas
particles in the same way as SNe feedback, i.e.\ smoothed over the
neighbour gas particles using the SPH smoothing algorithm
\citep[see Sec.\ 2.3.5 of][for details]{kg03a}.
Hence, we assume that the energy of the AGN heating 
is distributed spherically in the scale of 
the smoothing length ($\sim$ kpc) as purely thermal energy.
This is obviously too simple, and required to be improved.
However, we adopt this simple model for our first study.

We examine both the resulting X-ray and optical properties of the
simulation end-products, comparing them quantitatively with observation.
The gas particles in our simulations carry with them knowledge of the
density, temperature, and abundances of various heavy elements. Using the
XSPEC {\tt vmekal} plasma model, 
we derive
the X-ray spectrum for each gas particle, and synthesise them within the
assumed aperture.
We next generate
``fake'' spectra with the response function of the XMM EPN detector,
assuming an exposure time (40~ks) and target galaxy distance (17~Mpc). 
Finally, our XPSEC fitting provides the X-ray weighted temperatures and
abundances of various elements. 
 In the next section, we compare the simulation results with
the X-ray observational data in \citet[][MOM00]{mom00}. MOM00 used
the aperture radius of four times the $B$-band effective radius.
The $B$-band surface brightness profile for our successful model, 
i.e.\ Model 2, is barely able to be fitted by the Sersic 
($r^{1/n}$) law \citep{js68}\footnote{As mentioned in KG03b, 
the $B$-band surface brightness profile for Model 1, i.e.\ Model C in KG03b,
is too much centrally concentrated due to the continuous star formation
at the centre, and cannot be fitted by the Sersic law.},
which provides $R_{{\rm e},B}\approx11$ kpc.
Hence we applied a projected aperture radius of 
$R=45\approx 4 R_{{\rm e},B}$ kpc for both models.
Conversely, the simulated star particles
each carry their own age and metallicity ``tag'', which enables us to
generate an optical-to-near infrared spectral energy distribution for the
target galaxy, when combined with our population synthesis code 
\citep[which
itself is based upon the population synthesis models of][]{ka97}. 

\section{Discussion and Conclusions} 
\label{res-sec}

\begin{figure}
\centering
\includegraphics[width=\hsize]{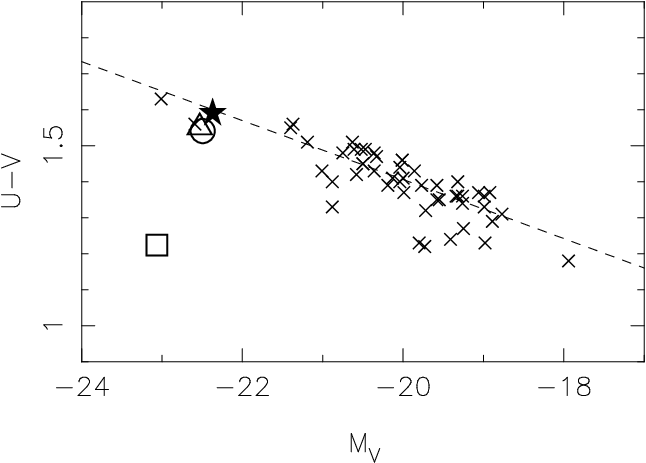}
\caption{
Comparison of the simulated $U-V$ CMR (square/circle/triagle
for Models 1/2/2L, respectively) and that of the Coma cluster ellipticals
(crosses). The observational data are from \citet{ble92a}. The dashed line
shows the CMR fitted to the Coma Cluster galaxies. For comparison, the solid 
star corresponds to NGC~3923
whose total magnitude is obtained from 
\citet{gdv91}, and colours are from \citet{pfa79}
who presented the colours within a aperture similar 
in size to what \citet{ble92a}
used.
}
\label{cmr-fig}
\end{figure}

To study both the X-ray and optical properties, we examine the ${\rm
L_X-T_X}$ and ${\rm L_X-[Fe/H]_X}$ relations for the X-ray properties, and
the colour-magnitude relation (CMR) for the optical properties, following
KG03b. Throughout this paper, we normalise all abundances to the solar
``meteoritic'' values from \citet{ag89}. Figs.\ \ref{lx-fig} and
\ref{cmr-fig} show the ${\rm L_X-T_X}$ and ${\rm L_X-[Fe/H]_X}$ relations
and the CMRs for Models 1 and 2 at $z=0$, compared that observed.  
Recall, Model~1 
corresponds to Model~C of KG03b; such models are broadly consistent with 
the empirical X-ray properties of ellipticals and agreement driven 
primarily by radiative cooling \citep[see
also][]{ffo96,ptce00,mtkpc01,bgw02,dkw02,rv03,tbm04}. 

\begin{figure}
\centering
\includegraphics[width=\hsize]{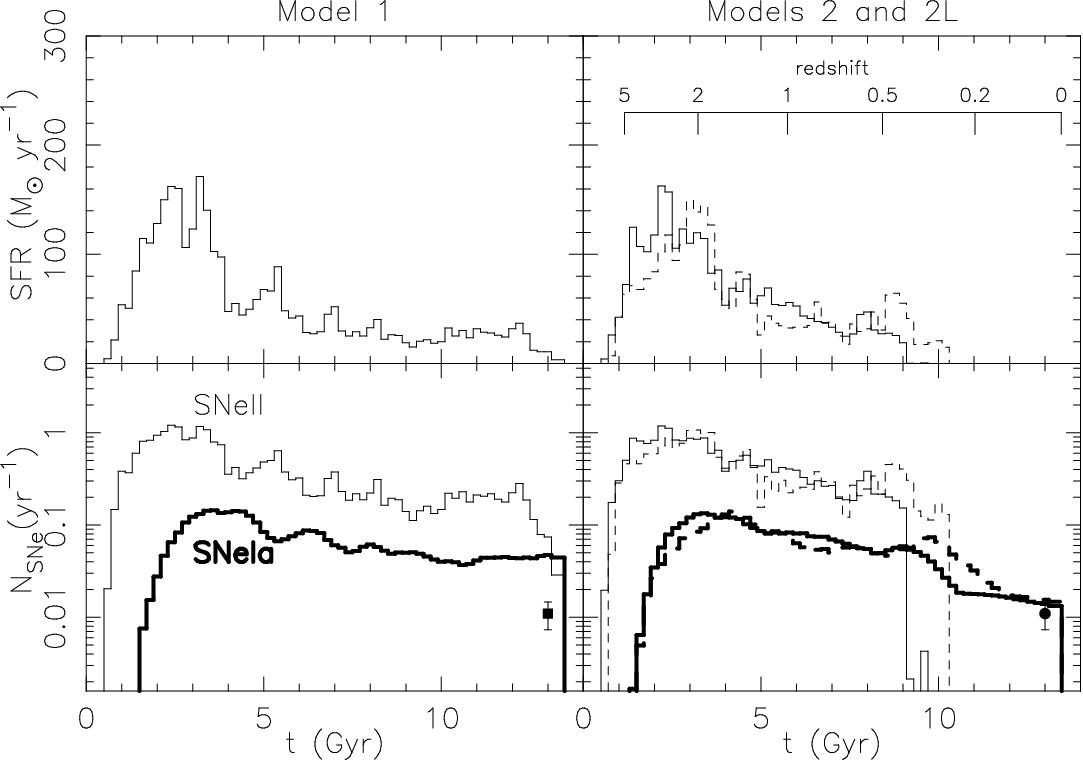}
\caption{
Time variation of the star formation rate (upper) and the event rate of
SNe II (thin lines) and SNe Ia (thick line) for Models 1 (left panels), 
2 (solid lines in right panels), and 2L (dashed lines in right panels).
Solid square (circle) with error-bar in lower left (right) panel
is taken from the observational SNe Ia rate by \citet{cet99},
taking into account the $B$-band luminosity for Model 1 (2).
To show them clearly, those SNe Ia rates are plotted at ${\rm t}=13$ Gyr. 
}
\label{sfr-fig}
\end{figure}

KG03b showed that radiative cooling ensures that the hot dense gas
turns into the cold (i.e.\ non-X-ray-emitting) gas, and keeps
X-ray-emitting gas high temperature and low density.
KG03b also found that 
stars preferentially enrich the gas in
the central region where radiative cooling is efficient. The high-density
cold gas is incorporated into future generations of stars,
and thus a large fraction of the iron ejected from stars is locked
into future generations of stars.
As a result, the hot gaseous halo which emits X-rays
is not enriched efficiently, leading to the lower [Fe/H]$_{\rm X}$ for
Model 1.
Hence, radiative cooling ensures that the hot gas has a lower
metallicity than the stellar component. However, the optical colours of
the resulting stellar component for this model are inconsistent with the
observational data (Fig.\ \ref{cmr-fig}), due to the excessive star
formation at low redshift (Fig.\ \ref{sfr-fig}) being driven by radiative
cooling (see KG03b for more details). 

In contrast, the new AGN heating model (Model 2) roughly reproduces both
the X-ray and optical observational data.
The shaded area in Fig.\ \ref{lx-fig} displays the range of ${\rm L_X}$
and ${\rm T_X}$ for a sample of elliptical galaxies with low-${\rm L_X}$ and
high-${\rm T_X}$ (group 5 in MOM00), and 90 \% confidence limit
of ${\rm [Fe/H]_X}$ from the composite X-ray spectrum of these galaxies.
The X-ray properties for Model 2 are consistent with 
those for the sample ellipticals.
In addition, Fig.\ \ref{cmr-fig} demonstrates that the colour and
magnitude for Model 2 are consistent with the CMR of the Coma cluster
ellipticals and roughly reproduce those for
NGC~3923\footnote{NGC~3923 is known
as a shell elliptical. However, their X-ray properties seem to be
normal, and old stellar population is dominant \citep[e.g.][]{mp04}.
Note that the effective radius of NGC~3923 ($R_{{\rm e},B}\sim$6.6 kpc)
is smaller than that for Model 2, which might affect the comparison
of the X-ray properties, because we adopted the aperture radius
of $4 R_{{\rm e},B}$. 
}
which is the brightest galaxy in this sample.

Of greatest importance for the analysis here is the fact that this AGN
heating is sufficiently efficient to suppress the late-time star formation
which plagued our earlier study (KG03b).  Fig.\ \ref{sfr-fig} demonstrates
this graphically. As a result, Model 2 succeeds in reproducing - \it
simultaneously \rm - both the X-ray and optical 
properties\footnote{Fig.\ \ref{sfr-fig} shows
that Model~2 also reproduces the Type~Ia SNe rate observed in nearby
ellipticals \citep{cet99} within the associated observational
uncertainties.}
observed in nearby ellipticals such as NGC~3923.
It is also worth noting that the lower ${\rm [Fe/H]_X}$ in Model 2 is 
induced by a different mechanism, i.e.\ blowing out the enriched gas, 
from that in Model 1, i.e.\ radiative cooling, because the iron ejected 
from stars is no long able to be hidden in the cold gas or
future generations of stars in Model 2.
In fact, we found that compared with Model 1, 
a larger fraction of iron ejected from stars 
is blown out from the system by the AGN heating.

\begin{figure}
\centering
\includegraphics[width=\hsize]{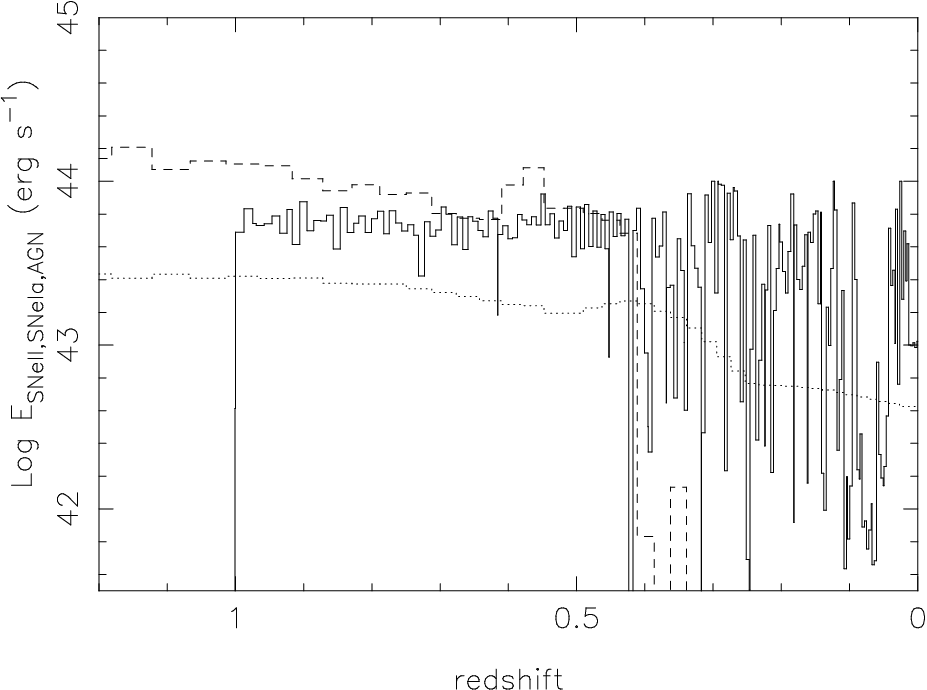}
\caption{
Histories of the feedback energy from Type~II SNe (dashed), Type~Ia SNe
(dotted) and the AGN (solid), for Model~2.
}
\label{fdeh-fig}
\end{figure}

Fig.\ \ref{fdeh-fig} shows the histories of the feedback energy from
SNe~II, SNe~Ia and the AGN.  Until star formation ceases at
$z$$\approx$0.4, the heating due to SNe~II is comparable to that of the
AGN. However, once star formation stops, the number of SNe~II drops, and
consequently they become unimportant as a heat source. Conversely, the AGN
provides continuous heat to the gas, as the AGN activity is not directly
linked to the star formation well outside the nucleus.  SNe~Ia heating by
itself is too inefficient to prevent the gas from cooling and subsequent
star formation occurring in Model 1. 
The mean heating energies from SNe~Ia and the
AGN over the final 3~Gyr of the simulation are $4.8\times10^{42}$ and
$2.5\times10^{43}$ ergs, respectively. The efficient, star
formation-independent, heating provided by the AGB is required to suppress
late-time star formation and, consequently, recover the red stellar
populations of ellipticals. 

Even after star formation ceases at $z$$\approx$0.4, the AGN activity becomes
more sporadic, because the gas fueling is also suppressed by the AGN
heating itself.\footnote{Note that our assumed AGN heating energy is
$10^{44}$ erg s$^{-1}$. Hence, in Fig.\ \ref{fdeh-fig}, AGN heating energy
of less than $10^{44}$ erg s$^{-1}$ means that the AGN activity has been
``switched on and off'' within the period of the size of that bin. This is
admittedly unrealistic, as the bin size is small (typically $\sim40$ Myr),
compared to the typical lifetime of the AGN activity ($\sim$100 Myr). 
This is a current limitation of the phenomenological AGN model employed.}
This self-regulation of the AGN activity ensures that the hot gas present
is not entirely removed from the system due to over-heating, or entirely
converted to cold gas by efficient radiative cooling.
For comparison, we 
also ran a model in which the AGN was continuously active (from redshift
$z$=1 to the present-day) independent of gas kinematics in the central
region.  We found that such continuous heating blows out the gas from the
system completely because of the efficient (and uninterrupted) heating. 

\begin{figure}
\centering
\includegraphics[width=\hsize]{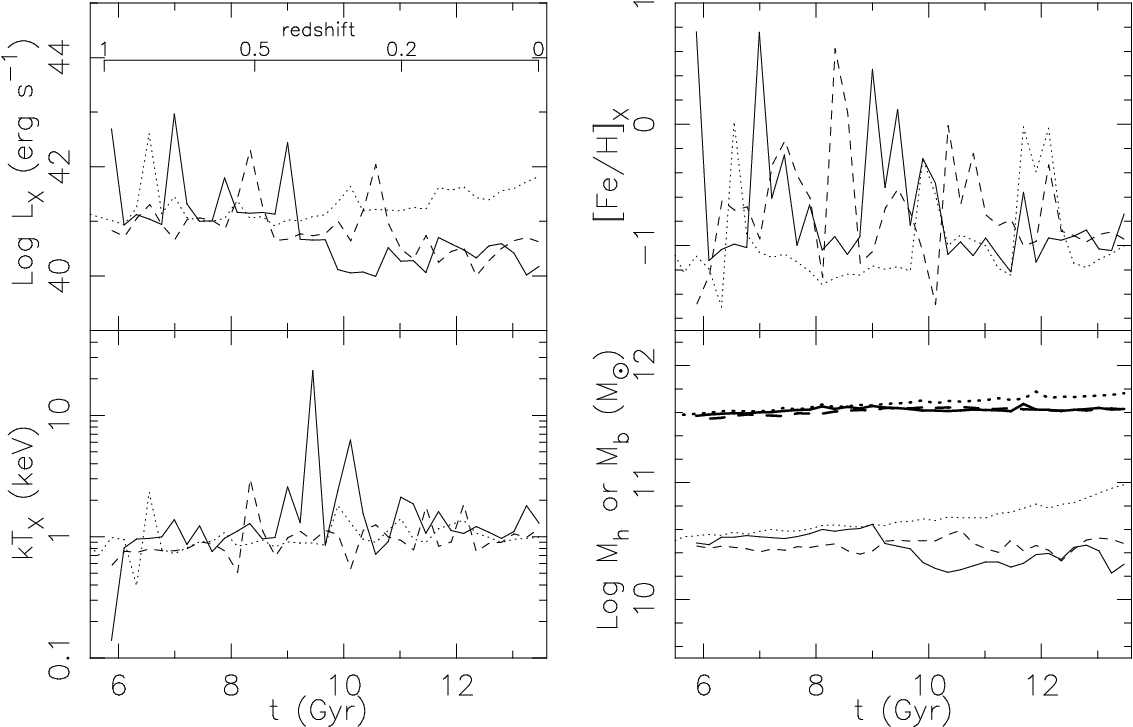}
\caption{
Time variation of the X-ray luminosity (upper left), temperature (lower
left), iron abundance (upper right) and the hot (thin lines) and total
baryon (thick line) mass (lower right) within $R=45$ kpc. The values
obtained in Models 1, 2, and 2L are presented as dotted, solid, and 
dashed lines, respectively. 
}
\label{lxtfemh-fig}
\end{figure}

Fig.\ \ref{lxtfemh-fig} displays the time variation of the relevant X-ray
properties for our simulated elliptical galaxies.
 Here, ${\rm T_X}$ and ${\rm [Fe/H]_X}$ are calculated 
by taking the mean values
weighted by $\rho^2 T^{1/2}$ for the gas with $T>10^{6}$ K,
following Fig.\ 12 of KG03b.
 Prior to the cessation
of star formation (at $z$$\approx$0.4), the X-ray properties for Model~2
vary sporadically, but settle into relevant stability thereafter.  We
suspect that this may be traced to our (too simple) assumption that the
AGN becomes active instantaneously at $z$=1; recall that this heating
source was introduced when the simulation was dynamically stable and in
possession of a deep potential well.  As such, there was a finite time
($\sim$2--3~Gyr) before the AGN could contribute sufficiently to the heat
of the surrounding ambient ISM and a new stable configuration reached.  In
reality, of course, it is likely that the AGN develops before the host
system becomes stable, and the AGN heating could affect the hot gas more
efficiently. In that picture, we can anticipate the hot gas properties
evolving more passively prior at $z$$>$0.4, in contrast with the current
models. Unfortunately, at this time, it is difficult to construct a more
self-consistent model using three-dimensional cosmological hydrodynamical
simulations. Nevertheless, our simple model demonstrates that the AGN
heating induced by gas infall leads to a stable condition for a lengthy
period of time due to its self-regulated nature. Hence, our results shown
in this paper are not ephemeral, but represent stable, long-term
properties of the simulated ellipticals. 

 After the hot gas properties reached a stable condition,
the X-ray properties are still varying a little. This temporal variance
is considered to represent the range of uncertainties in our
simulation\footnote{Some features like a periodic change of ${\rm L_X}$
in Model 2 
seem to be real, because it is caused by the expansion and re-collapse
of the hot gas due to the self-regulated AGN activity. However, we postpone
such discussion until higher resolution simulations become available.}. 
Thus, Fig.\ \ref{lxtfemh-fig} means that our simulation cannot
tell any difference in ${\rm T_X}$ or ${\rm [Fe/H]_X}$ between
Models 1 and 2. Conversely, the difference in ${\rm L_X}$ is significant.
The range of ${\rm Log(L_X)}$ since $z=0.2$ ($t\approx11$ Gyr) is
41.2$-$41.8 and 40.0$-$40.7 for Models 1 and 2, respectively.
Consequently, the AGN heating model reduces the X-ray luminosity for
the target galaxy by at least a factor of five, 
and leads to the X-ray properties similar to
low-${\rm L_X}$ and high-${\rm T_X}$ ellipticals . It would be interesting
to examine whether AGN heating can produce the other type of elliptical
galaxies, such as high-${\rm L_X}$ ellipticals, because it seems that
there is no difference in optical properties between low-${\rm L_X}$
and high-${\rm L_X}$ ellipticals \citep{km00}.

 Finally, to demonstrate the uncertainty by the numerical resolution,
we carried out a simulation (Model 2L) in which the high-resolution region 
(Sec.\ \ref{meth-sec}) consists of 
lower-resolution particles ($27/64$ by mass), and the same
parameters as Model 2 are adopted.
Note that Model 2L has a different pattern of density fluctuations
in the high-resolution region, which leads to a different minor merger
history from Model 2.
Fig.\ \ref{lx-fig} shows that
Model 2L provides similar X-ray properties to Model 2 within
the range of the temporal variance seen in Fig.\ \ref{lxtfemh-fig}, i.e.\
this variance corresponds to the uncertainties of our simulation
as described above.
Star formation is quenched by the AGN heating at $z\sim0.3$
(Fig.\ \ref{sfr-fig}), and the red colour is reproduced (Fig.\ \ref{cmr-fig}).
Hence, our conclusion shown above is less
sensitive to the numerical resolution.

Our cosmological simulation with the AGN heating has successfully produced
a system which is consistent with observed elliptical galaxies in both the
X-ray and optical regimes. To our knowledge, this is the first elliptical
galaxy model which does so \it self-consistently\rm.  We admit that our
AGN model is rather simple and our study limited by its focus upon only
one (representative) elliptical galaxy simulation.  
Nevertheless, we believe that this encouraging result 
provides additional impetus for galactic astronomy community to
further explore the role that AGN heating plays in the formation and
evolution of elliptical galaxies.

\section*{Acknowledgements}

The financial support of the Australian Research Council, through
its Discovery Project scheme, is gratefully acknowledged.
We acknowledge the Yukawa Institute Computer Facility,
the Astronomical Data Analysis Center of the National Astronomical
Observatory, Japan (project ID: rmn12a), the
Institute of Space and Astronautical Science 
of Japan Aerospace Exploration Agency, and
the Australian and Victorian Partnerships for Advanced
Computing, where the numerical computations for this paper were
performed. This research was also supported in part by the
National Science Foundation under Grant No. PHY99-07949.

\label{lastpage}


\begin{thebibliography}{999}
\small
\bibitem[Allen et al.\ (2001)]{atn01}
 Allen S.W.\ et al., 2001, MNRAS, 324, 842
\bibitem[Anders \& Grevesse (1989)]{ag89}
 Anders E., Grevesse N., 1989, Geochim. Cosmochim. Acta, 53, 197
\bibitem[Basson \& Alexander (2003)]{ba03}
 Basson J.F., Alexander P., 2003, MNRAS, 339, 353
\bibitem[Bertschinger (2001)]{eb01}
 Bertschinger E., 2002, ApJS, 137, 1
\bibitem[Binney \& Tabor (1995)]{bt95}
 Binney J.J.\ \& Tabor G.\ 1995, MNRAS 276, 663
\bibitem[Blanton, Sarazin \& McNamara (2003)]{bsm03}
 Blanton E.L., Sarazin C.L., McNamara B.R.\ 2003, ApJ 585, 227
\bibitem[Blanton et al.\ (2001)]{bsmw01}
 Blanton E.L., Sarazin C.L., McNamaraB.R., Wise M.W., 2001,
 ApJL, 558, L15
\bibitem[Borgani et al.\ (2002)]{bgw02}
 Borgani S.\ et al., 2002, MNRAS 336, 409 
\bibitem[Borgani et al.\ (2004)]{bms04}
 Borgani S.\ et al., 2004, MNRAS, 348, 1078
\bibitem[Bower, Lucey \& Ellis (1992)]{ble92a}
 Bower R.G., Lucey J.R., Ellis R.S., 1992, MNRAS, 254, 589
\bibitem[B\"ohringer et al.\ (1993)]{bvfen93}
 B\"ohringer H., Voges W., Fabian A.C., Edge A.C., Neumann D.M.,
 1993, MNRAS, 264, L25
\bibitem[Brighenti \& Mathews (1999)]{bm99}
 Brighenti F., Matthews W.G., 1999, ApJ, 515, 542 
\bibitem[Brighenti \& Mathews (2002)]{bm02}
 Brighenti F.\ \& Mathews W.G.\ 2002 ApJL, 574, L11
\bibitem[Br\"uggen \& Kaiser (2002)]{bk02}
 Br\"uggen M., Kaiser C.R., 2002, Nature, 418, 301
\bibitem[Caperrallo, Evans \& Turatto (1999)]{cet99}
 Cappellaro E., Evans R., Turatto M., 1999, A\&A, 351, 459
\bibitem[Churazov et al.\ (2001)]{cbkbf01}
 Churazov E., Br\"uggen M., Kaiser C.R., B\"ohringer H., Forman W.,
 2001, ApJ, 554, 261
\bibitem[Churazov et al.\ (2002)]{csfb02}
 Churazov E., Sunyaev R.,  Forman W., B\"ohringer H.,
 2002, MNRAS, 332, 729
\bibitem[Ciotti \& Ostriker (2001)]{co01}
 Ciotti L., Ostriker J.P., ApJ, 2001, 551, 131
\bibitem[Dav\'e, Katz \& Weinberg (2002)]{dkw02}
 Dav\'e R., Katz N., Weinberg D.H., 2002, ApJ, 579, 23
\bibitem[de Vaucouleurs et al.\ (1991)]{gdv91}
 de Vaucouleurs G., de Vaucouleurs A., Corwin H. G., Buta R. J., 
 Paturel G., Fouque P., 1991, Third Reference Catalogue of Bright
 Galaxies (New York: Springer-Verlag)
\bibitem[Finoguenov \& Jones (2001)]{fj01}
 Finoguenov A., Jones C., 2001, ApJL, 547, L107
\bibitem[Fujita, Fukumoto \& Okoshi (1996)]{ffo96}
 Fujita Y., Fukumoto J., Okoshi K., 1996, ApJ, 470, 762
\bibitem[Fujita et al.\ (2002)]{fsk02}
 Fujita Y.\ et al., 2002, ApJ, 575, 764
\bibitem[Granato et al.\ (2004)]{gds04}
 Granato G.L., De Zotti G., Silva L., Bressan A., Danese L., 
 2004, ApJ, 600, 580
\bibitem[Heintz et al.\ (2002)]{hcrb02}
 Heintz S., Choi Y.-Y.,  Reynolds C.S., Begelman M.C., 2002, ApJL, 569, L79
\bibitem[Hoeft \& Br\"ueggen (2004)]{hb04}
 Hoeft M., Br\"uggen M., submitted to ApJ (astro-ph/0405434)
\bibitem[Jones et al.\ (2002)]{jfv02}
 Jones C.\ et al., 2002, ApJL, 567, L115
\bibitem[Kawata \& Gibson (2003a)]{kg03a} 
 Kawata D., Gibson B.K., MNRAS, 2003, 340, 908
\bibitem[Kawata \& Gibson (2003b)]{kg03b}
 Kawata D.\ \& Gibson B.K.\ 2003b, MNRAS 346, 135 (KG03b)
\bibitem[Kay, Thomas \& Theuns (2003)]{ktt03}
 Kay S.T., Thomas P.A., Theuns T., MNRAS, 2003, 343, 608
\bibitem[Kodama \& Arimoto (1997)]{ka97} 
 Kodama T., Arimoto N., 1997, A\&A, 320, 41
\bibitem[Kodama \& Matsushita (2000)]{km00} 
 Kodama T., Matsushita K., ApJ, 539, 149
\bibitem[Lewis et al.\ (2000)]{lbk00}
 Lewis G.F., Babul, A., Katz N., Quinn T., Hernquist L.,
 Weinberg D.H., 2000, ApJ, 536, 623
\bibitem[Matsushita, Ohashi \& Makishima (2000)]{mom00}
 Matsushita K., Ohashi T., Makishima K., 2000, PASJ, 52, 685 (MOM00)
\bibitem[McNamara et al.\ (2000)]{mwn00}
 McNamara B.R.\ et al., 2000, ApJL, 534, L135
\bibitem[Meza et al.\ (2003)]{mnse03}
 Meza A., Navarro J.F., Steinmetz M., Eke V., 2003, ApJ, 590, 619
\bibitem[Michard \& Prugniel (2004)]{mp04}
 Michard R., Prugniel P., A\&A, 2004, 423, 833
\bibitem[Muanwong et al.~(2001)]{mtkpc01}
 Muanwong O., Thomas P.A., Kay S.T., Pearce F.R., Couchman H.M.P.,
 2001, ApJL, 552, L27
\bibitem[Nulsen (2004)]{pn04}
 Nulsen P., 2004, in Reiprich T., Kempner J., Soker N., eds,
 Proc.\ The Riddle of Cooling Flows in Galaxies and Cluster of Galaxies,
 http://www.astro.virginia.edu.au/coolflow/  
\bibitem[Omma et al.\ (2004)]{obbs04}
 Omma H., Binney J., Bryan G., Slyz A., MNRAS, 2004, 348, 1105
\bibitem[Pearce et al.~(2000)]{ptce00}
 Pearce F.R., Thomas P.A.,  Couchman M.P., Edge A.C., 2000, MNRAS,
 317, 1029
\bibitem[Persson, Frogel \& Aaronson (1979)]{pfa79}
 Persson S.E., Frogel J.A., Aaronson M., 1979, ApJS, 39, 61
\bibitem[Quilis, Bower \& Balogh (2001)]{qbb01}
 Quilis V., Bower R.G., Balogh M.L., 2001, MNRAS, 2001, 1091
\bibitem[Saxton, Bicknell \& Sutherland (2002)]{sbs02}
 Saxton C.J., Bicknell G.V., Sutherland R.S., 2002, ApJ, 579, 176
\bibitem[Scannapieco \& Oh (2004)]{so04}
 Scannapieco E., Oh P.S., 2004, ApJ, 608, 62
\bibitem[Sersic (1968)]{js68}
 Sersic J.-L., 1968, Atlas de Galaxieas Australes
 (Cordoba: Observatorio Astronomico)
\bibitem[Suginohara \& Ostriker (1998)]{so98}
 Suginohara T., Ostriker J.P., 1998, ApJ, 507, 16
\bibitem[Springel, Di matteo, \& Hernquist (2004a)]{sdh04a}
 Springel V., Di Matteo T., Hernquist L., 2004b, ApJL submitted
 (astro-ph/0409436)
\bibitem[Springel, Di matteo, \& Hernquist (2004b)]{sdh04b}
 Springel V., Di Matteo T., Hernquist L., 2004b, MNRAS submitted
 (astro-ph/0411108)
\bibitem[Tornatore et al.\ (2004)]{tbm04}
 Tornatore L., Borgani S., Matteucci F., Recchi S., Tozzi P.,
 2004, MNRAS, 349, 19
\bibitem[Tornatore et al.\ (2003)]{tbs03}
 Tornatore L.\ et al., 2003, MNRAS, 342, 1025
\bibitem[Valdarnini (2003)]{rv03}
 Valdarnini R., 2003, MNRAS, 339, 1117
\end{thebibliography}
\end{document}